# Tripod of Requirements in Horizontal Heterogeneous Mobile Cloud Computing

ZOHREH SANAEI[1], SAEID ABOLFAZLI[2], ABDULLAH GANI[3], RASHID HAFEEZ KHOKHAR[4]
Mobile Cloud Computing Research Lab, Faculty of Computer Science and Information Technology
University of Malaya
MALAYSIA
sanaei[1], abolfazli[2]@ieee.org, abdullah[3], rashid[4]@um.edu.my

*Abstract:* - Recent trend of mobile computing is emerging toward executing resource-intensive applications in mobile devices regardless of underlying resource restrictions (e.g. limited processor and energy) that necessitate imminent technologies. Prosperity of cloud computing in stationary computers breeds Mobile Cloud Computing (MCC) technology that aims to augment computing and storage capabilities of mobile devices besides conserving energy. However, MCC is more heterogeneous and unreliable (due to wireless connectivity) compare to cloud computing. Problems like variations in OS, data fragmentation, and security and privacy discourage and decelerate implementation and pervasiveness of MCC. In this paper, we describe MCC as a horizontal heterogeneous ecosystem and identify thirteen critical metrics and approaches that influence on mobile-cloud solutions and success of MCC. We divide them into three major classes, namely ubiquity, trust, and energy efficiency and devise a tripod of requirements in MCC. Our proposed tripod shows that success of MCC is achievable by reducing mobility challenges (e.g. seamless connectivity, fragmentation), increasing trust, and enhancing energy efficiency.

## 1 Introduction

Smartphones are rapidly growing and their increasing popularity is expected to surpass stationary computing devices [1]. However, their proliferation toward fulfilling ever-increasing outlooks of end-user is decelerated by their restrained resources, especially unreplenishable energy.

Employing cloud computing technology in mobile computing created Mobile Cloud Computing (MCC) that is aimed with multiple visions: computing augmentation, storage extension, cost reduction, battery conservation, and data safety enhancement. Abundant opportunities of cloud is expected to enhance computing experience of nearly 240 million mobile business users in few years and drive in revenue of more than US$5 billion [2].

Moreover next generation wireless networks [3] is deemed to provide opportunities of continuous, consistent mobile connectivity, high data-rate transmission, controlled jitter, reduced delay, and support heterogeneous environments via access technologies like WLAN and 3G. However, leveraging MCC technology is hindered by several challenges like application and data fragmentation, security and privacy, jitter, communication delay, and signal handoff [9]. In addition, energy efficiency of mobile augmentation approaches in this domain is of importance.

Addressing these challenges and problems is important to unleash power of MCC that demands certain key requirements; the absence of such requirements degrades quality and leads to uncertainty of mobile users, developers, mobile operators, and cloud providers. In this paper, we extract the term horizontal heterogeneity in MCC domain using qualitative approach and analyze critical requirements and approaches toward ideal MCC. Accordingly, we categorize them into three main groups of ubiquity, trust, and energy efficiency and propose tripod model of requirements in this ecosystem. This tripod would be a scale to evaluate usefulness of cloud services, mobile-cloud solutions and approaches.

Rest of the paper is organized as follow: in Section 2 the horizontal heterogeneity in mobile cloud computing is discussed from three aspects of cloud, mobile, and network. Analysis of key requirements is presented in Section 3 and paper is concluded in Section 4.

## 2 Horizontal Heterogeneity

MCC is an amalgam of three foundations of cloud computing, mobile computing, and networking. Large variation in platforms causes horizontal heterogeneity and complicates MCC more than cloud computing. The term horizontal heterogeneity indicates magnificent difference between each aspect. Such classification provides better insight to the concept and facilitates analysis of key requirements.

### 2.1 Cloud Computing

Cloud computing provides multiple services of Infrastructure as a service (IaaS), Platform as a Service (PaaS), and Software as a Service (SaaS). Numerous cloud vendors provide services, platforms, and APIs that intensify heterogeneity of cloud landscape and create interoperability [4] as the major challenge in cloud computing exacerbated with mobility.

In addition, business competition diversifies cloud providers with proprietary frameworks that increase heterogeneity in cloud environment [5]. For instance, if an application uses IaaS and is deployed in Structured Query Language (SQL) based server, it is almost impossible to be ported to PaaS cloud based on Google Query Language (GQL) due to difference in implementation techniques. Insisting on porting these applications levies upfront investment and consideration [6].

Moving among single type of cloud like IaaS is less challenging by leveraging open virtualization format [7] that allows cloud-users deploy their virtual appliances at any cloud provider.

### 2.2 Mobile Computing

In mobile ecosystem, growing variation in device specification, architecture, operating system, programming language, and API create application and data fragmentation. It requires great deal of efforts to execute a BlackBerry application on Nokia handset. In addition, different flavors of an operating system offer unique features and services that are not compatible with other versions. For instance, from android family, HoneyComb_MR2 offers 13-level API while HoneyComb_MR1 supports 12- level API. Unlike PCs, smartphones are not upgradable and user cannot benefit from technology advancement by updating and spending nominal amount.

### 2.3 Networking

Connectivity is a distinctive metric between MCC and cloud computing. Wireless technologies composition makes MCC more complicated than wired communication. Intermittent wireless medium consists of various networking technologies such as WLAN, 3G, and WiMax degrades quality of connectivity that is worsening by device mobility. Variation in network technologies affects on mobile application responsiveness due to dissimilar characteristics of engaging networks in global roaming. When a mobile user enters low-bandwidth network area, drastic drop in data transmission lowers quality of application response in case of heavy traffic.

Moreover, low security of wireless network and probability of signal interception are challenges which are expected to be increased by crossing the channel of Internet in order to maintain cloud services for mobile augmentation.

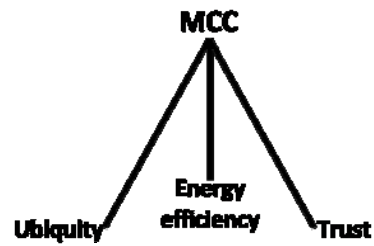

Fig. 1: Tripod of MCC Requirement

## 3 Vital Tripod of MCC

MCC aims to augment multitude of weak mobile devices by leveraging heterogeneous domain of cloud through various wireless technologies. Several proposals seek to alleviate smartphone resource deficiency by leveraging cloud computing such as MAUI [8] which are suffering from challenges like fragmentation, security, and privacy on MCC.

We identify thirteen critical metrics and approaches that influence on MCC and mobile-cloud solutions and classify them under three main classes of ubiquity, trust, and energy efficiency. Addressing these challenges, leads to success of MCC and mobile-cloud approaches. Figure 1 depicts our proposed tripod.

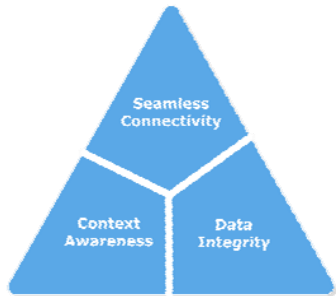

Fig. 2: Ubiquity Essentials.

## 3.1 Ubiquity

Ubiquity in MCC is not confined to the movement ability. It is state of the art online mobility with support of integrity and interoperability in context-aware ecosystem that is achievable in convergence of three phenomena as shown in Figure 2.

### 3.1.1 Seamless connectivity

Establishing and maintaining reliable connection between nomadic device and different entities in wireless medium is a critical issue to fully unleash power of MCC. Having seamless connectivity between heterogeneous wireless networks in MCC necessitates reliable inter-system handoff schemes along next generation wireless networks [9]. Next generation wireless networks with IP-based infrastructure are envisioned to provide seamless connectivity across heterogeneous wireless networks to avoid network blips and interruptions that are escalated via mobility. However, connotation of IP-based architecture proposals may convey always-on connectivity, high data rate services, and finally enhanced quality of user experience, but it requires an intelligent system to manage mobility in global roaming which is yet under investigation.

### 3.1.2 Data Integrity

Cloud providers envision providing context-agnostic distributed resources to end-users and enabling them to access services and data anytime anywhere regardless of underlying device and platform. Data fragmentation and vendor lock-in problems hinder interoperability in clouds that needs to be addressed. In addition, a mechanism is required to ensure user that data modification is stored and not exposed or modified (either by cloud provider or third party) since the last access. In well-integrated domain, mobile application retrieves and process distributed data from silo of various clouds.

### 3.1.3 Context Awareness

Wireless connectivity, power interruption (dead battery), heterogeneity, mobility as well as users profile and social situation are accumulating to make MCC a rapidly changing ecosystem. Metrics like bandwidth variation, intermittent connectivity, service heterogeneity, and user emotion impact on quality of experience and obliges accurate awareness from the surrounding environment. Thereby, context awareness becomes inseparable phenomenon in MCC.

## 3.2 Energy Efficiency

Energy is the only unrestorable resource in mobile devices which cannot be replenished without external resource [3] and current technologies are able to increase battery capacity 5% per annum only [10]. Several energy harvesting efforts since the 1990's aim to replenish energy from resources like human movement [11] and wireless radiation [12] that still are under investigation. Alternatively, several methods are proposed to conserve mobile's local resources and reduce the resource hunger of mobile applications that are discussed as follows and are depicted in Figure 3.

### 3.2.1 Remote Execution

An early effort to conserve energy in mobile devices is cyber-foraging [13] that introduces the notion of conserving native resource of mobile device by migrating resource hungry part(s) of application to the nearby resource-rich computing machine which is connected to uninterruptable power source and internet. Though there are several efforts in literature like [8, 14] but still it is under investigation for acceptable performance [15].

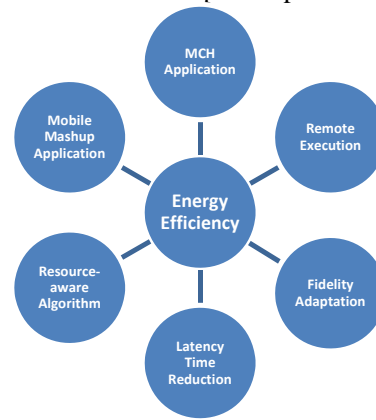

Fig. 3: Energy Efficiency Approaches in MCC.

### 3.2.2 Fidelity Adaptation
Another approach that is potentially useful in conserving device's resources is fidelity adaptation (FA) which trades-off execution quality with energy conservation [16]. Energy is conserved and application can be executed by altering application-level quality metrics to exchange quality of service with local resources [17].

### 3.2.3 Latency Time Reduction
Latency negatively impacts on energy efficiency [18] and interactive response [19] especially in wireless network where latency varies from one technology to another. Also, in cellular communication, distance to/from base station (near or far) effects on energy efficiency of data transmission.

Bandwidth and speed variation of wireless technologies limit their usability. For example data transfer bit-rate imposes comparatively more impact on energy efficiency of cellular networks than WLAN. The higher the transmission bit-rate, the more energy efficient the transmission [18].

Moreover, leveraging wireless Internet network to execute process-intensive applications on distant rich resources creates bottleneck plus long WAN latency that degrades quality of user experience. Cloudlet [20] is an effort to mitigate interaction latency of mobile users by deploying WLAN instead of HSPDA (High Speed Download Packet Access) to access remote resources.

### 3.2.4 Resource-aware Algorithm
Considering limited resource of mobile devices, creating resource-aware mobile applications and technologies reduces considerable amount of resource consumption. Various power reduction techniques are orchestrated to deliver best quality computing with least power consumption. There are several energy-aware memory management studies like [21, 22] to reduce the energy consumption of storing data. Mobile Ram and Phase Change Memory (PCM) [23, 24] are common memory technologies used in mobile phones. In Mobile Ram, power management unit maintains multiple power states like 'Self Refresh' and 'Power Down' to minimize power consumption whereas PCM leverages three states of I/O, on, and off to store data with enhanced energy consumption [25].

### 3.2.5 Mobile-Cloud Hybrid (MCH) Applications
MCH applications aim to minimize the resource requirement and consumption of an application without quality trade-off. Cloud resources run the application quickly and mitigated communication overhead reduces energy consumption and overall execution time.

$\mu$Cloud [26] is a mobile application framework leveraging cloud. Three types of components namely mobile, cloud, and hybrid are presented to develop mobile applications with least component-level communication and dependency with promise of high functionality and resource efficiency. In order to reduce resource consumption, and energy resource-intensive components are located inside the cloud to be executed at runtime and push the result to mobile device.

### 3.2.6 Mobile Mashup Application
Mobile Mashup (MM) can be leveraged to reduce energy and resource requirement of mobile applications. MM is a technique to create mobile application by aggregating available *services* and *contents* offered in ubiquitous environment leveraging Mobile Service Oriented Architecture (Mobile SOA) [27] as the backbone. During runtime, application calls service execution externally and collects and reintegrates response.

However, mashup using SOA is a complex approach, but several efforts seek to justify and fit it into mobile ecosystem to benefit from loosely coupling and remote execution of services toward conserving local resources like [28-30] to name few approaches toward green mobile computing.

## 3.3 Trust
Trust is a metric by which user can evaluate consequences of other's action on his own activities in a collaborative domain. Mobile users, cloud providers, and third parties (can be client of cloud user or service broker) are concerned with trust [31, 32]. Fundamentally trust is a personal trait. Pessimism and optimism, nature and sensitivity of request, and service provider reputation, interfere on level of trust between client and cloud.

We identify five metrics that are required at trust establishment phase to be considered by trustee and truster which are depicted in Figure 4 and explained as follows:

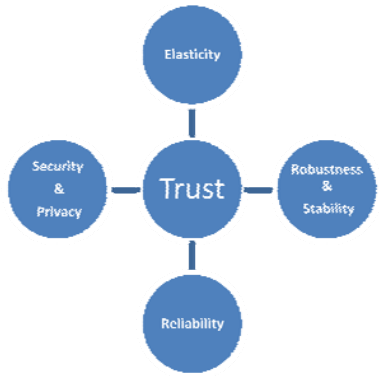

Fig. 4: Trust Metrics in MCC

### 3.3.1 Elasticity
Cloud provider also confronts situations that requests are more than available resources that obliges resource provisioning. Ability of on-demand resource provisioning without service interruption is an important property of trustworthy cloud provider which ensures that increase in load does not decrease quality.

### 3.3.2 Reliability
Clients do not have control over the data and code execution inside the cloud. Thus clients become uncertain to receive what they pay for. Service-Level Agreement (SLA) is leveraged to confine cloud providers to a certain level of QoS and ensures clients to receive what is paid for. According to user requirements and service provider capability, SLA is a scale to enforce and monitor QoS and assures service availability and responsiveness accordingly.

### 3.3.3 Robustness and Stability
Being public resource exposes cloud to fault and failure. Robust architecture and fault tolerance of cloud is a key to enhance level of trust among users. Stability guaranties unique functionality and correctness of output in multiple service request iterations.

### 3.3.4 Security and Privacy
In first glimpse, security is visible issue for requesters only. However, in order to maintain a secure runtime and storage environment, service providers also require assurance that users will not violate terms of service; malware execution is potentially a risk with catastrophic consequences for service provider. Intrinsically, trust is not necessarily mutual. If client trusts cloud, it does not guaranty that cloud also trusts client and vice versa. Hence, trust is influenced by quality of security and privacy mechanisms implemented in MCC environment.

## 4 Conclusion
Mobile cloud computing is state of the art technology of mobile computing leveraged cloud computing to conserve energy and augment processing as well as storage ability of mobile devices in wireless environment. Inward dissimilarity of mobile devices' implementation and cloud servers, inhomogeneity and limited bandwidth on wireless technologies compare to fixed networks make MCC more challenging than cloud computing. Distant resources, dynamically changing environment because of user mobility, data fragmentation, wide diversity of applications, and wireless links are sample problems in MCC success and growth route. These problems bring the challenges like interoperability, security and privacy, and seamless connectivity which need to be addressed. Implementing ideal MCC as a multidisciplinary technology stands on three cornerstones: ubiquity, energy efficiency, and trust. In order to support these cornerstones, primary approaches (e.g. fidelity adaptation, mobile cloud hybrid applications, and context awareness) and metrics (e.g. elasticity, robustness and stability, and latency time reduction) are required. Accumulation of cornerstones, approaches, and metrics is potentially suitable for evaluating usefulness and pervasiveness of MCC service and MCC-based solutions.

## Acknowledgment
This work is fully funded by Malaysian Ministry of Higher Education under the University of Malaya High Impact Research Grant UM.C/HIR/MOHE/FCSIT/03.

*References:*
[1] Gartner Lowers PC Forecast as Consumers Diversify Computing Needs Across Devices, Acceesible from: www.gartner.com/it/page.jsp?id= 1570714, 2011.
[2] Mobile Cloud Application, ABI Research Report, Acceesible from: www.abiresearch.com /research/1003385.

[3] I. F. AKYILDIZ, et al., A Survey of Mobility Management in Next-Generation All-IP-Based Wireless Systems, IEEE Wireless Communications, 2004.

[4] M. Hogan, et al., NIST Cloud Computing Standards Roadmap, July 2011.

[5] D.DURKEE, Why Cloud Computing Will Never Be Free, Communication of the ACM, Vol.53, 2010, pp. 62-69.

[6] A. Ranabahu and A. Sheth, Semantics Centric Solutions for Application and Data Portability in Cloud Computing, Cloud Computing Technology and Science, IEEE Second International Conference 2010, pp. 234 - 241.

[7] OVF, Open Virtualization Format Specification, Acceesible from: www.dmtf.org/sites/default/files/standards/documents/DSP0243_1.0.0.pdf, 2009.

[8] E. Cuerv, et al., Maui: Making Smartphones Last Longer with Code Offload, MobiSys:Proc. of the 8th international conference on Mobile systems, applications, and services, 2010.

[9] N. Nasser, et al.,Handoffs in Fourth Generation Heterogeneous Networks, Communications Magazine, Vol.44, No.10, 2006, pp. 96-103.

[10] M. Satyanarayanan, Avoiding Dead Batteries, Pervasive Computing, Vol.4, 2005, pp. 2-3.

[11] S. Robinson, Cellphone Energy Gap: Desperately Seeking Solutions, Strategy Analytics, Acceesible from: www.strategyanalytics.com/default.aspx?mod=reportabstractviewer&a0=4645, 2009.

[12] T. Starner, et al., The Locust Swarm: An Environmentally-Powered, Networkless Location and Messaging System, 1st International Symposium on Wearable Computers Digest of Papers.,1997, pp.169-170.

[13] M. Satyanarayanan, Pervasive Computing: Vision and Challenges, Personal Communications,Vol.8, No.4, 2001, pp. 10-17.

[14] S. Kosta, et al., Unleashing the Power of Mobile Cloud Computing Using Thinkair, Arxiv preprint arXiv: 1105.3232, 2011.

[15] M. Sharifi, et al.,A Survey and Taxonomy of Cyber Foraging of Mobile Devices, Communications Surveys & Tutorials, IEEE, Vol.PP, No.99, 2012, pp. 1-12.

[16] B. Rajesh Krishna, Powerful Change Part 2: Reducing the Power Demands of Mobile Devices, Pervasive Computing, IEEE, Vol.3, No.2, 2004, pp. 71-73.

[17] B. D. Noble, et al., Agile Application-Aware Adaptation for Mobility, Proc.of the sixteenth ACM symposium on operating systems principles, Vol.31, 1997, pp. 276-287.

[18] A.P.Miettinen and J.K.Nurminen, Energy Efficiency of Mobile Clients in Cloud Computing, HotCloud'10 Proc. of the 2nd USENIX conference on Hot topics in cloud computing, 2010, pp. 4-4.

[19] H. Lagar-Cavilla, et al., Interactive Resource-Intensive Applications Made Easy, Proc. 8th International Middleware Conference, 2007, pp. 143-163.

[20] M. Satyanarayanan, et al., The Case for Vm-Based Cloudlets in Mobile Computing, Pervasive Computing, IEEE, 2009.

[21] E. De Lara, et al., Puppeteer: Component-Based Adaptation for Mobile Computing, Proc. 3rd USENIX Symposium on Internet Technologies and Systems, 2001, pp. 14-25.

[22] J. Flinn and M. Satyanarayanan, Energy-Aware Adaptation for Mobile Applications, Proc. of the seventeenth ACM symposium on Operating systems principles, Vol.33, 1999, pp. 48-63.

[23] M. Lee, et al., PABC: Power-Aware Buffer Cache Management for Low Power Consumption, Computers, IEEE Transactions on, Vol.56, No.4, 2007, pp. 488-501.

[24] H. Huang, et al., Design and Implementation of Power-Aware Virtual Memory, Proc. of the annual conference on USENIX Annual Technical 2003.

[25] B. G. Johnson and C. H. Dennison, Patent No.: 6,791,102, 2004.

[26] R. Duan, et al., Exploring Memory Energy Optimizations in Smartphones, Internationa Conference and Workshops on Green Computing 2011, pp. 1-8.

[27] L. Yan, et al., Virtualized Screen: A Third Element for Cloud Mobile Convergence, Multimedia, Vol.18, No.2, 2011, pp. 4-11.

[28] R. Tergujeff, et al., Mobile Soa: Service Orientation on Lightweight Mobile Devices, IEEE International Conference on Web Services, 2007, pp. 1224-1225.

[29] Y. Natchetoi, et al., Service-Oriented Architecture for Mobile Applications Proc. of the 1st International workshop on Software architectures and mobility, 2008.

[30] M. Aoyama, et al., Attribute-Based Architecture Patterns for Lightweight Service-Oriented Architectures, Software Engineering Conference 2009, pp. 119-126.

[31] A. IM and A. Martin, A Trust in the Cloud, Information Security Technical Report, 2011.

[32] Smartphone/Tablet User Survay, Jun 2011, Prosper Mobile Insight Acceesible from: www.prospermobileinsights.com/Default.aspx?pg=19.